\documentclass[a4paper,11pt]{article}
\usepackage[utf8]{inputenc}
\usepackage{graphicx}
\usepackage{amsmath}
\usepackage{amssymb}
\usepackage{latexsym}
\usepackage{subcaption}
\usepackage[colorlinks]{hyperref}
\usepackage{color}
\usepackage{enumerate}
\usepackage{enumitem}
\usepackage{cite}
\usepackage{makeidx}
\usepackage{colortbl}
\title{\bf Low Energy Consequences of Loop Quantum Gravity}
\author{{\bf Salman Sajad Wani}$^1$, {\bf Behnam Pourhassan}$^2$ {\bf Mir Faizal}$^{3,4}$ and {\bf Ahmed Jellal}$^5$ \\
$^1${\it Department of Physics,
University of Kashmir,  Kashmir, India} \\
$^2${\it School of Physics, Damghan University, Damghan, 3671641167, Iran} \\
$^3${\it Department of Physics and Astronomy,
University of Lethbridge, Lethbridge, AB T1K 3M4, Canada} \\
$^4${\it Irving K. Barber School of Arts and Sciences,  University of British
Columbia--Okanagan,}  \\  {\it Kelowna, British Columbia V1V 1V7, Canada}\\
$^5${\it Laboratory of Theoretical Physics,  
Faculty of Sciences, Choua\"ib Doukkali University,}\\
{\it  PO Box 20, 24000 El Jadida, Morocco
}}
\date{}
\begin{document}

\maketitle

 \begin{abstract}
As the loop quantum gravity is based on polymer quantization,
  we will  argue that the polymer length (like string length)
 can be several orders larger than the Planck length,
 and this can  have low energy consequences.   We  will
 demonstrate that a short distance modification of a quantum system by polymer
 quantization and by string theoretical considerations can produce similar behavior. Moreover,  it will be demonstrated that  a family
 of different deformed Heisenberg algebras can produce similar low energy effects.  We will analyze such  polymer correction
 to a degenerate Fermi gases in a harmonic trap, and its   polymer corrected  thermodynamics.
\end{abstract}

 The polymer quantization
has been motivated from loop quantum gravity  \cite{n1, n2}. In fact,
  the polymer quantization of simple matter fields has been studied
  and the
conclusions reached resemble  the conclusions
obtained in loop quantum gravity \cite{p1, p2}. It has been demonstrated that the polymer quantization leads to
radiation in inertial frames \cite{p4}, 
and  a modification of   black body radiation    at high temperatures
\cite{p5}. The usually polymer length in polymer quantization is assumed to be of the same order as the Planck length.
However, it is possible for the polymer length to be
larger then Planck length,  and this can have low energy consequences (which can be experimentally measured).
It may be noted that it has already  been argued that the minimal length can be related to the
string length $l_s$  as   
 $l_{min} = g_s^{1/4} l_s$,
 with  $g_s$ being  the string coupling constant  \cite{s16}. In fact, the short distance correction to
 the  quantum mechanics from  a string length much greater than Planck length  has been analyzed \cite{ml12, ml21}.
 It has been suggested that  Landau levels and Lamb shift would be modified by
  a deformation of quantum mechanics produced from such a minimal length, and this can be measured experimentally  \cite{ml15}.
It has also been proposed that an  opto-mechanical setup can be used to measure such a deformation, which occurs
at several orders of magnitude larger than the Planck scale \cite{ml14}.

Motivated by the mentioned  results above in string theory,
we consider the polymer length to be several orders of magnitude larger than Planck length (only    bounded by
experiments). 
We will show that this can modify short distance
behavior of quantum systems. Indeed, we will analyze the corrections brought
by the polymer quantization to the Fermi-Dirac distribution
as well as to the thermodynamics of a Fermi gas in harmonic trap.

We start by analyzing the effect polymer quantization would have on the
 Heisenberg algebra. It is known that in the polymer quantization,  the usual phase space
$(x, p)$ can be  replaced by  \cite{p1, p2}
\begin{eqnarray}
 x \longrightarrow x, && p \longrightarrow \frac{1}{\mu} \sin (\mu p)
\end{eqnarray}
where $\mu$ is the polymer length scale. This mapping modifies
the usual Heisenberg algebra  to
\begin{eqnarray}\label{fpp}
 [x, p] \longrightarrow \frac{1}{\mu} [x, \sin (\mu p)]
 =  i \left(1 - \frac{\mu^2 p^2}{2}+f(p)\right)
\end{eqnarray}
where the function of momentum $f(p) = \mu^{4} p^{4} + \cdots$. Now it is known that such a deformation of the Heisenberg algebra will produce non-trivial modification to various physical systems, even if we neglect the higher order corrections given by $f(p)$ (and consider corrections of the order  $p^2$) \cite{ml12}-\cite{1}.   Note that
the first order correction 
to  the ordinary  Heisenberg
algebra 
 is of order $\mu^2 p^2$, which  
 is the correction that will be observed in any low energy future experiment.
These
corrections (of the form $\mu^2 p^2$) are similar to those obtained from generalized uncertainty principle
\cite{1,2}. Such  modification of uncertainty principle  is motivated from  string theory and has similar consequences
as polymer quantization \cite{1a}.
As in string theory, fundamental strings are the smallest probes, then it is not possible to probe spacetime below
string length scale, which leads to a string length being the minimal measurable length in spacetime \cite{z2}.
This modification of spacetime by minimal measurable length of the order of string length $l_s = \alpha'$
gives rise to
a generalized uncertainty principle
\cite{cscds,2z}. In perturbative string theory, the minimal measurable length is related to the string length $l_s$  as
 $l_{min} = g_s^{1/4} l_s$   \cite{s16}.
Even though non-perturbative effects add point like probes to string theory, such as $D0$-branes,
it has been demonstrated that
even with such non-perturbative objects, a minimal length of the order
$l_{min} = l_s g_s^{1/3}$  exists in string theory  \cite{s16, s18}.
As      the winding number
$w$ and the excitation number $n$ can be used to obtain the total energy of quantized strings,
it can be argued that the description of string theory
below   $l_s$   is the same as its description   above $l_s$ \cite{s16}.
This is because under $T$-duality,  we have the correspondences   $R\longrightarrow l_s^2/R$, and
$n\longrightarrow w$.   The $T$-duality has also been used to study the behavior of
effective path integral
for string   propagating in compactified extra dimensions, and similar conclusions have been obtained
form such  system   \cite{green1, green2}. The construction of
double field theories is   motivated from $T$-duality \cite{df12, df14}, and it has been observed that
a minimal length also exist in   double field theories \cite{mi15}. Thus, such  deformation of Heisenberg
algebra can  be motivated from string theory results.

Now, as polymer quantization, and the modification
of quantum mechanics by string theory, lead to the same first order
 corrections to the Heisenberg algebra, it can be argued that like string length scale, polymer length scale can also be several orders
 of magnitude larger than the Planck scale. In fact, it is possible that in nature two polymer scales exist, such that one of them is the
 gravitational polymer scale $\mu_{gr}$, and the other is the polymer scale for matter fields $\mu$ (as polymer quantization of matter fields has also been
 studied \cite{p1, p2}), such that  $\mu_{gr} = g \mu $, where $g$ is some coupling in a more fundamental theory.
 Under this similarity, it is possible
 to analyze physical consequences of polymer quantization on condensed matter systems in similar way
 to those  of generalized uncertainty
principle on such low energy system \cite{ml12, ml21, ml15, ml14}.
It may be noted that  any  deformation of the Heisenberg algebra  $
 [x, p] =  i \left(1 - \frac{\mu^2 p^2}{2}+f(p)\cdots\right).
$
with any function of
$f(p) = \mu^{4} p^{4} + \cdots$,
cannot be   distinguishable from the polymer quantization,  as any near future experiment would first
measure the first order correction of the form $\mu^2 p^2$.
As the non-local   deformation of the Heisenberg algebra,
\begin{equation}
 [x, p]= i\sqrt{1 - \mu^2 p^2}
\end{equation}
cannot be distinguished by near future experiments from polymer quantization,   it  has
been used to analyze the consequences of polymer quantization \cite{7}.
It is also possible to take other   modifications of Heisenberg algebra, such as
 \begin{equation}
  [x, p]= i\cos \mu p
 \end{equation}
 and all these  can be distinguished from polymer quantization
by any near future experiments, because all these first order corrections are of the form $\mu^2 p^2$.

Now we can study the effect of such short distance corrections on a degenerate Fermi gas in  a harmonic trap,
as such a system can be physically realized   \cite{k1, k2, k4, k5}. In fact,
  spin-polarized fermionic alkali atoms, like $^{6}Li$ and $^{40}K$, have  been studied  in  magnetic traps  \cite{be12, be14, be15, be16}.
This has motivated the study of  Fermi gas in isotropic
or anisotropic harmonic traps  \cite{A, A1}. In this study, the  thermodynamic    properties of such
Fermi gas have been analyzed the  trapping of  ultracold
atoms in a harmonic potential. It is expected that such a system will be very
sensitive to short distance modification of quantum system. So, a modification of such thermodynamic properties
for such a system can be experimentally measured \cite{a, a1}.
The  Fermi polarons have been studied using  spin down impurities in a spin up Fermi sea of ultracold atoms \cite{a2}.
The  polaron energy   for various interaction strengths have been studied around a Feshbach resonance, for this system.
The investigation of  Fermi polarons in a spin-imbalanced Fermi gas has been done using
  using  photoemission spectroscopy \cite{q1, q2}.
Now as these systems would be modified by polymer quantization,
it is important to understand the deformation of such a system by polymer quantization, as this can be measured
experimentally.

It is expected that a deformation of the Heisenberg algebra will deform the angular momentum algebra. Now as
the first order consequences of polymer quantization, and generalized uncertainty principle have the same
mathematical form, we can use the deformation of the angular momentum algebra by generalized uncertainty principle
\cite{saur} to write the polymer  deform   algebra of angular momentum as
\begin{equation}
 L^{2} \vert l,m\rangle = \hbar^{2}l(l+1){\left(1-\mu^{2} p^{2}\right)}^{2}\vert l,m\rangle, \qquad
 L_{z}\vert l,m\rangle=\hbar m\left(1-\mu^{2} p^{2}\right) \vert l,m\rangle.
\end{equation}
As the  Fermi gas in isotropic
or anisotropic harmonic traps  \cite{A, A1} have been studied, and can be measured experimentally,
we can analyze the polymer quantization of the
following isotropic harmonic potential
\begin{equation}
 V(x,y,z)=\frac{1}{2}M\omega^2 \left( x^2+y^2+z^2 \right)
\end{equation}
and the  polymer  corrections to the  energy eigenvalues can be expressed as
\begin{equation}
 E_n=\omega\left(n+\frac{3}{2} \right) \left(1-\mu^{2} p^{2} \right)
\end{equation}
where the
degeneracy of shell with energy  $ E_n $ is  given by
\begin{equation}
 g_n=\frac{1}{2}(n+1)(n+2).
\end{equation}
The total number  of quantum
states $ S_\alpha $  with energy less or equal to $E_\alpha $  can be written as
\begin{equation}
 \sum_{n=0}^\alpha g_n
=\frac{1}{6}(\alpha+1)(\alpha +2)(\alpha +3).
\end{equation}

\begin{figure}[h!]
 \begin{center}$
 \begin{array}{cccc}
\includegraphics[width=75 mm]{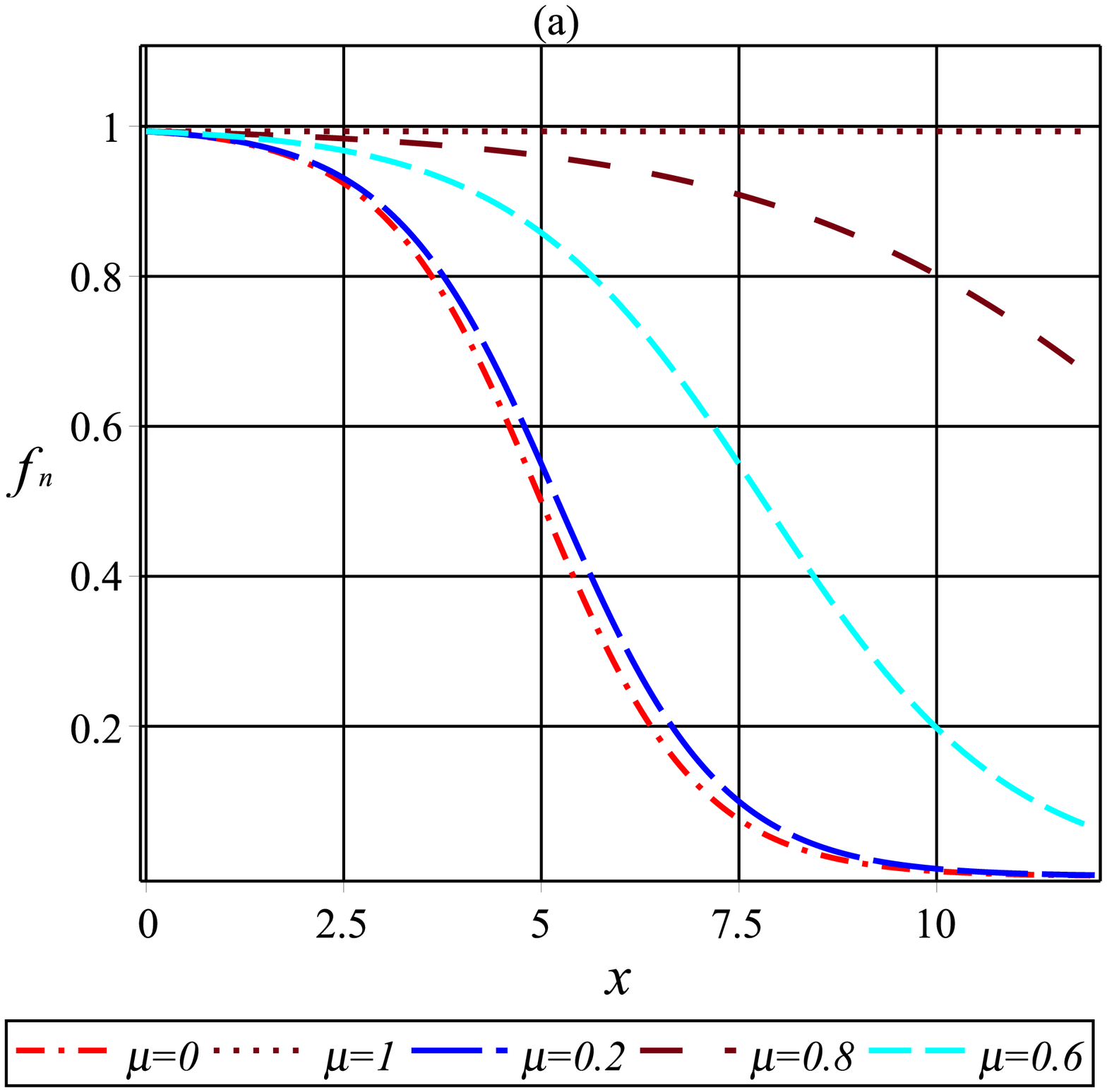}\includegraphics[width=75 mm]{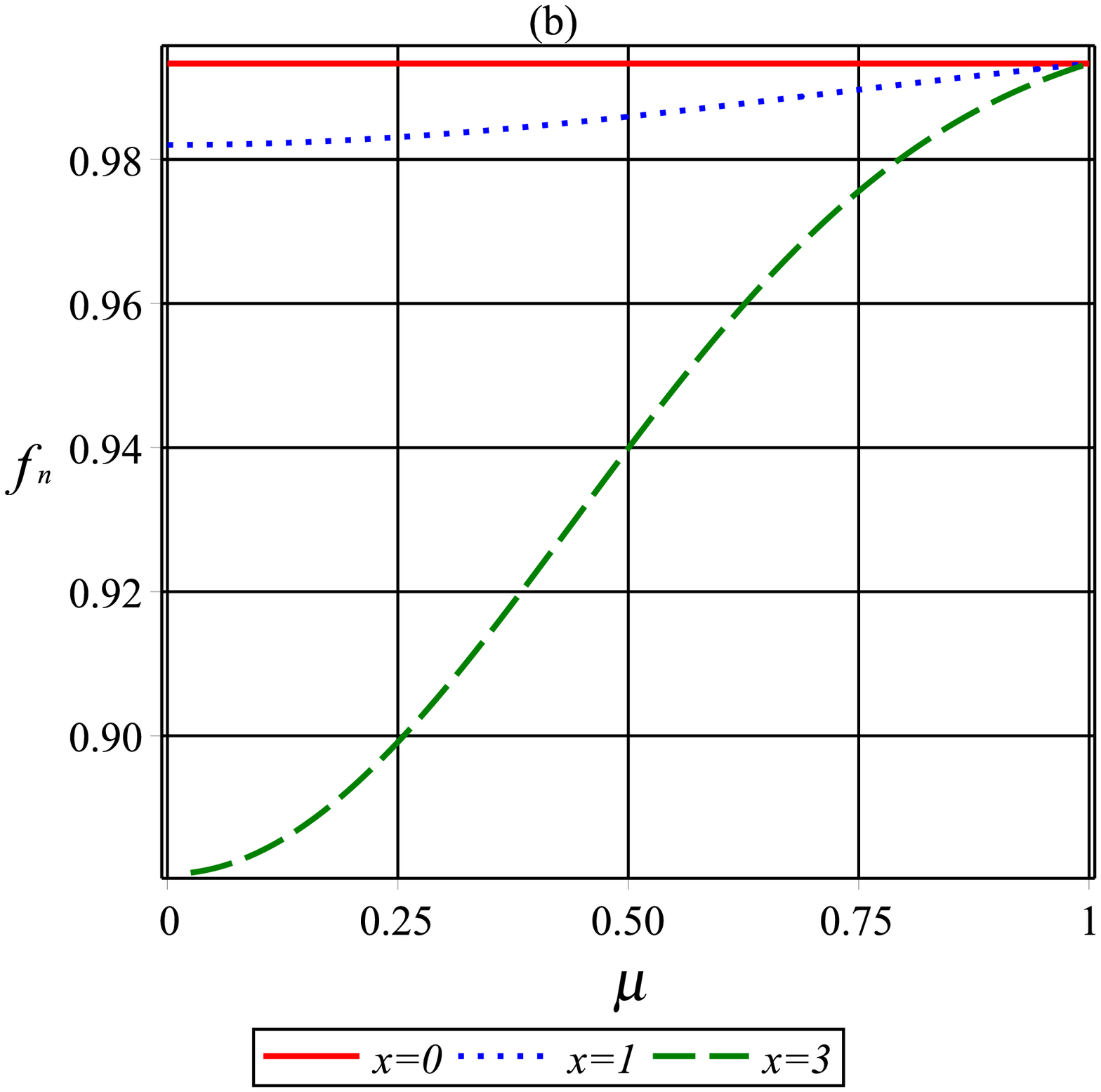}
 \end{array}$
 \end{center}
\caption{Fermi-Dirac distribution function for $p=1$.}
 \label{fig1}
\end{figure}

Now we can obtain the polymer deformation of
  Fermi-Dirac distribution
\begin{equation}
 f_n=\frac{1}{{e^{-\beta\nu +\beta\omega\left(n+\frac{3}{2}\right)(1-\mu^{2} p^{2})}+1}}
\end{equation}
where $\nu$ is the chemical potential
and  $\beta=\frac{1}{k_B T}$, with $k_B$ is the Boltzmann constant.
The total number of particles
in the trap and  Fermi energy are given by
\begin{equation}
 N=\sum_n g_nf_n, \qquad
E_F=\omega\left(n_F+\frac{3}{2}\right) \left(1-\mu^{2} p^{2}\right).
\end{equation}
 In the Fig. \ref{fig1}, we plot the deformed Fermi-Dirac distribution function $f_n$
 to observe the corrections due to the polymer quantization.
 We set $-\beta\nu\equiv y=5$, and $\beta\omega\left(n+\frac{3}{2}\right)\equiv x$, to plot
 $f_{n}$ in terms of $x$ for several values of $\mu$ in Fig. \ref{fig1} (a) and in terms of correction parameter for several values of $x$ in Fig. \ref{fig1} (b).
 In the case of $\mu=0$ (dash dotted red line of Fig. \ref{fig1} (a)),
 we have usual Fermi-Dirac distribution. 
 For $\mu <1$,
 $f_n$ increases (see Fig. \ref{fig1} (b)) and
the Fermi energy is reduced by $\mu$. 
For $\mu \geq1$ as well as $x=0$ (corresponding to high temperature), we can find $f_{n}\approx1$ as expected.

\begin{figure}[h!]
 \begin{center}$
 \begin{array}{cccc}
\includegraphics[width=90 mm]{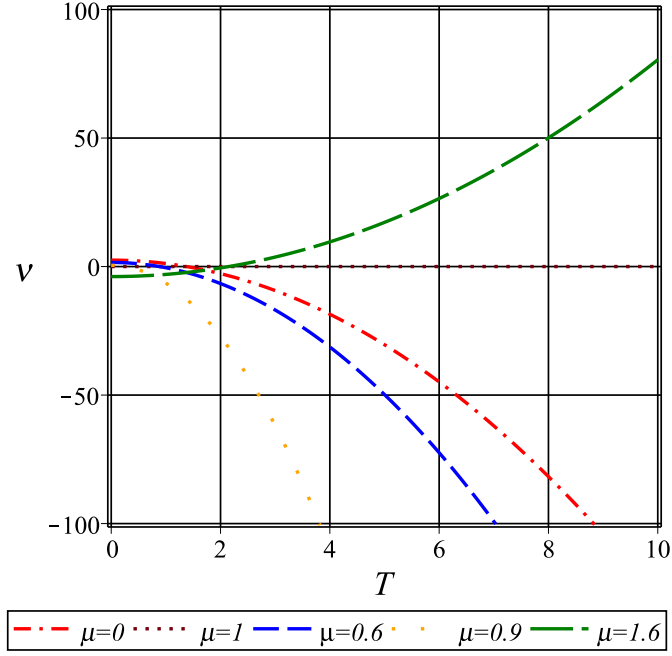}
 \end{array}$
 \end{center}
\caption{Typical behavior of chemical potential for $n_{F}=k_{B}=\omega=p1$.}
 \label{fig2}
\end{figure}

For high temperature regime,   the chemical potential can be written as
\begin{equation}\label{chem}
 \nu(T)=\omega \left(n_F+\frac{3}{2}\right) \left(1-\mu^{2} p^{2}\right)
  \left(1-\frac{\pi^2}{3}\left(\frac{k_BT}{\omega\left(n_F+\frac{3}{2}\right)\left(1-\mu^{2} p^{2}\right) }\right)^2 \right)
\end{equation}
In the Fig. \ref{fig2}, we can observe the typical behavior of the chemical potential for different values of $\mu p$.
We can observe a completely opposite behavior for   $\mu >1$ and $\mu <1$.
As expected, for the small correction ($\mu <1$), we the behavior resembles the  uncorrected chemical potential.
Both of them coincide at low temperature. Hence, the   polymer corrections are only important at high temperature.

However, in order to analyze the present system for low temperature regime, we start by first defining
 the number $N_>(T)$ of  exited atoms above $E_F$,  number $N_<(T)$ of atoms below $E_F$  and corrected  short
 distance,
 \begin{eqnarray}
 &&  N_>(T)=\sum_{n=n_{F}+1}^\infty \frac{g_n}{z^{-1}e^{-n \omega \beta(1-\mu^{2} p^{2})}+1}\label{nsup}\\
  && N_<(T)=\sum_{n=0}^\infty g_n \left(1-{\frac{1}{ z^{-1}e^{-n\omega \beta(1-\mu^{2} p^{2})}+1}} \right)\label{ninf}
 \end{eqnarray}
 where the fugacity is given by
 \begin{equation}
  z =e^{\beta(\nu-\frac{3}{2}\hbar\omega(1-\mu^{2} p^{2}) )} \\
 \end{equation}
For low temperature regime, we have two conditions, i.e.,
$ k_B T \ll (1-\mu^{2} p^{2}) E_{n_F+1}-\nu$ and  $k_B T \ll \nu -(1-\mu^{2} p^{2}) E_F$. Under these conditions
we can write 
 \begin{eqnarray}
  && \Sigma_>= \sum_{n=n_{F}+1}^\infty g_ne^{-(E_n-E_{n_{F}+1})(1-\mu^{2} p^{2})\beta}\\
  && \Sigma_< = \sum_{v=0}^{n_F} g_n e^{-(E_F-E_n)(1-\mu^{2} p^{2})\beta}
 \end{eqnarray}
Now we can use the following approximation,
\begin{eqnarray}
 && { N_>}(T)\approx \Sigma_>e^{-\beta((1-\mu^{2} p^{2}) E_{n_{F}+1}-\nu) }\\
 && N_<(T)\approx \Sigma_<e^{-\beta(\nu-(1-\mu^{2} p^{2}) E_{n_F})}
\end{eqnarray}
The product of $N_>(T)$ and $N_<(T)$ can now be written as
\begin{equation}
N_>(T)N_<(T)\approx\Sigma_>\Sigma_<e^{-\beta(1-\mu^{2} p^{2}) \omega}.
\end{equation}
So, when $N\in \{S_\alpha\}$, the Fermi shell is  totally filled at $T=0$, $N_>(T)=N_<(T)$, and hence
we obtain
\begin{equation}
 N_>(T)=\left(\Sigma_>\Sigma_<e^{-(1-\mu^{2} p^{2}) \beta\omega}\right)^{-\frac{1}{2}}
\end{equation}
Therefore the chemical potential takes the form
\begin{equation}
\nu(T)= \omega(n_F+2)(1-\mu^{2} p^{2})-\frac{1}{2}k_B T \ \ln  \frac{\Sigma_>}{\Sigma_<}.
\end{equation}
We can also analyze    the low temperature limit for   $N \not\in\{S_\alpha\}$.
For very low temperature regime,  the Fermi function can be approximated by $f_n=1$ for
$n < n_F$ and $f_n = 0$ for $n > n_F$. The number of occupied states
in the Fermi shell
$\bigtriangleup N=N-S_{n_{f}-1}$ can be approximated by
%
\begin{eqnarray}
 \bigtriangleup N
= \frac{g_n}{{z^{-1}e^{\beta n_F \omega(1-\mu^{2} p^{2}) }}+1}
\end{eqnarray}
and then
the corresponding chemical potential is given by,
\begin{equation}
 \nu(T)=\omega\left(n_f+\frac{3}{2}\right) \left(1-\mu^{2} p^{2}\right)
 -k_B T\ \ln\left({{\frac{g_{v_F}}{\bigtriangleup N}}}-1\right).
\end{equation}
 Finally, the   energy of the present  system can be written as
\begin{equation} \label{U}
U(T)=\sum_{n=0}^\infty \frac{g_n n\omega (1-\mu^{2} p^{2}) }{{z^{-1}e^{n\omega (1-\mu^{2} p^{2})\beta }}+1}
\end{equation}
We find that, effect of polymer correction is increasing of the energy. Thus, all the thermodynamic quantities for a harmonic oscillator
are deformed because of the polymer quantization.
Such modification to various important thermodynamic quantities can change the properties of such Fermi gases,
which
can in detected  experimentally.

\begin{figure}[h!]
 \begin{center}$
 \begin{array}{cccc}
\includegraphics[width=90 mm]{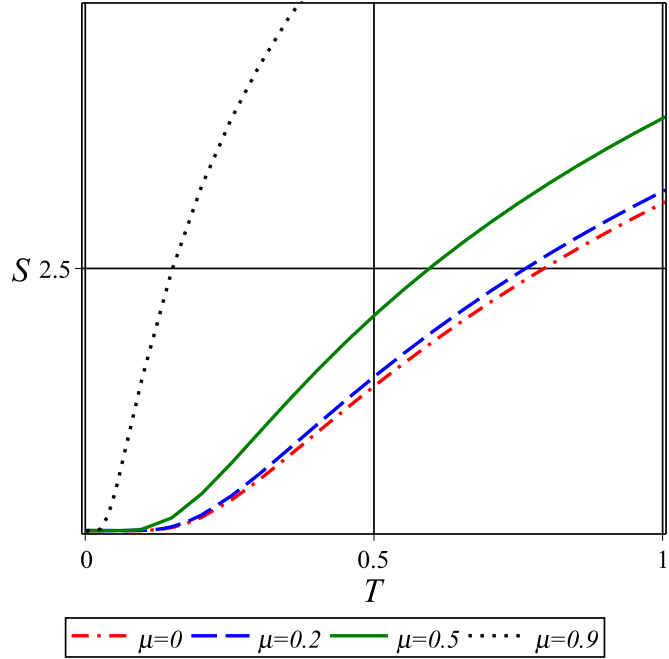}
 \end{array}$
 \end{center}
\caption{Typical behavior of the entropy for $p=k_{B}=\omega=1$.}
 \label{fig3}
\end{figure}

In that case the single grand partition function is given by
\begin{equation}
\mathcal{Z} =\text{Tr} e^{-\beta H}= \text{Tr} e^{\beta\nu-\beta\alpha\omega\left(n+\frac{3}{2}\right)}
\end{equation}
with $\alpha= (1-\mu^{2} p^{2})$. Now the  partition function can be written as
\begin{equation}
\mathcal{Z} =\frac{e^{\beta\nu}}{\left(2\sinh\frac{\beta\alpha \omega}{2}\right)^3}
\end{equation}
Hence, we can obtain entropy by using the following relation
\begin{equation}
S=k_{B}\ln\mathcal{Z}+k_{B}T\left(\frac{\partial\ln\mathcal{Z}}{\partial T}\right)_{V}.
\end{equation}
In the Fig. \ref{fig3}, we observe the  effect of correction on the entropy.
It is clear that polymer correction increases value of the entropy.

\begin{figure}[h!]
 \begin{center}$
 \begin{array}{cccc}
\includegraphics[width=90 mm]{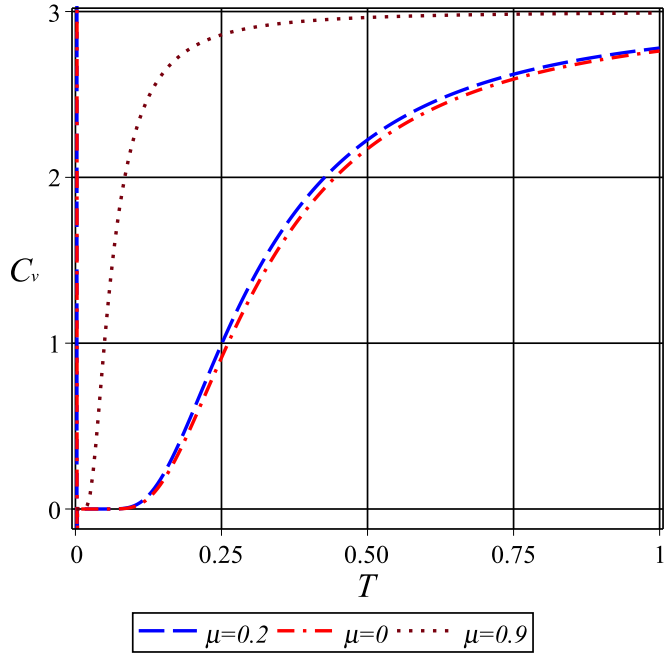}
 \end{array}$
 \end{center}
\caption{Typical behavior of the specific heat in constant volume for $p=k_{B}=\omega=1$.}
 \label{fig4}
\end{figure}

The specific heat in constant volume is obtained  using,
\begin{equation}
C_{V}=T\left(\frac{\partial S}{\partial T}\right)_{V}.
\end{equation}
In the plots of Fig. \ref{fig4}, we draw specific heat for some values of correction parameter.
For $\mu \geq1$, we find that specific heat
is negative.  This can be interpreted as instability in the system. For $\mu <1$,
the specific heat is increasing function of the temperature (see Fig. \ref{fig4}).
At low temperature the effect of $\mu$ is negligible and the the specific
heat tends to zero. For the high temperature limit it yields to a constant.

\begin{figure}[h!]
 \begin{center}$
 \begin{array}{cccc}
\includegraphics[width=90 mm]{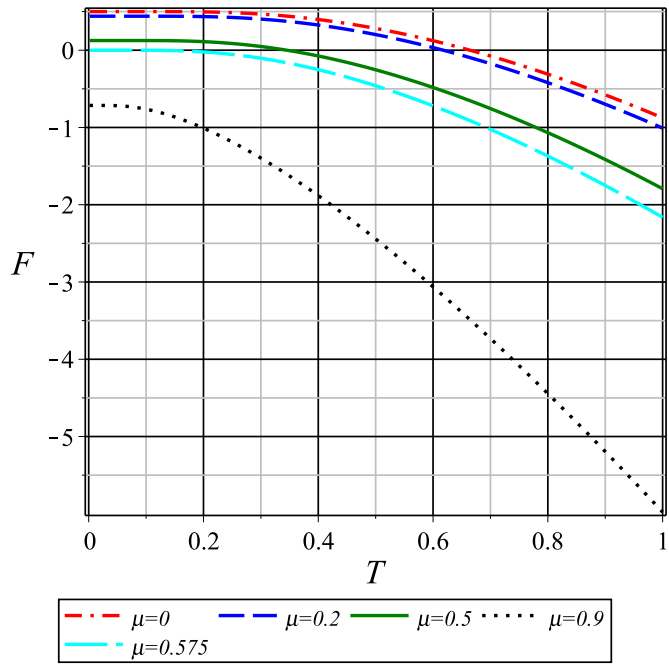}
 \end{array}$
 \end{center}
\caption{Typical behavior of the  Helmholtz free energy versus $T$ for $p=k_{B}=\omega=1$.}
 \label{figF}
\end{figure}

We can also calculate Helmholtz free energy as,
\begin{equation}
F=-k_{B}T\ln\mathcal{Z}.
\end{equation}
We find critical value for the correction parameter where Helmholtz free energy is zero (see long dash cyan line of Fig. \ref{figF}).
Below which Helmholtz free energy is positive at low temperature and yields to negative value at high temperature. Otherwise for the larger
value of $\mu$ value of Helmholtz free energy is totally negative. There is a maximum for the Helmholtz free energy which may be interpreted as unstable point.

\begin{figure}[h!]
 \begin{center}$
 \begin{array}{cccc}
\includegraphics[width=80 mm]{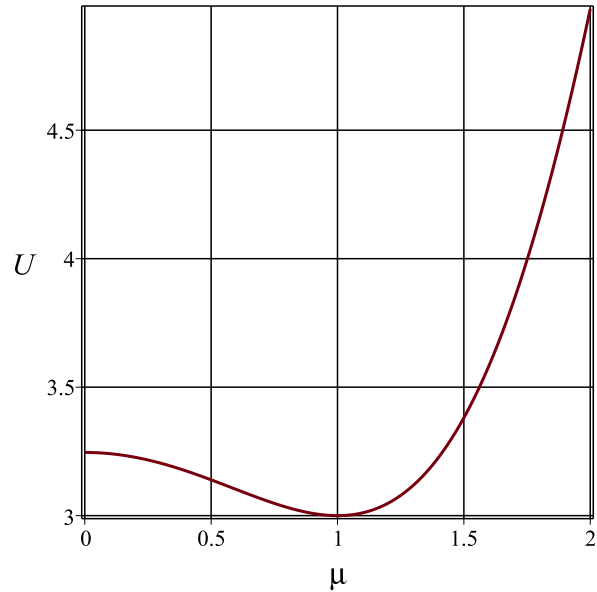}
 \end{array}$
 \end{center}
\caption{Typical behavior of the internal energy versus $\mu$ for $p=k_{B}=T=\omega=1$.}
 \label{fig5}
\end{figure}

Finally it is interesting to see behavior of the internal energy with correction parameter
which is illustrated by the Fig. \ref{fig5}.
We observe  that the internal energy given by the equation (\ref{U})
has a minimum for special value of $\mu$. For the case of $\mu \leq1$
we can see that internal energy is decreasing function of $\mu$.
Comparing this with the experimental value of internal energy could be used to obtain a bound on the  value of $\mu$.

Motivated by the deformation of Heisenberg algebra brought from string theory, and its low energy consequences
\cite{ml12, ml21, ml15, ml14},
 we argued that the polymer length
for matter fields can be much greater than Planck length. Such a modification of polymer quantization
can have interesting low energy consequences,
and these can  experimentally be measured. As an example, we analyzed
the generate Fermi gas in a harmonic trap using such deformation. We explicitly
analyzed the modification to Fermi-Dirac distribution as well as to the thermodynamics of the
present system.
It has been demonstrated that polymer quantization leads to a loss of manifest Lorentz symmetry in Minkowski space.
\cite{p2}. So, it would be interesting to analyze the
  effect of polymer quantization on the Poincare algebra \cite{IJTP, CSF}. This can have important consequences for
  particle physics, if the polymer length is assumed to be much larger than Planck scale. It would be interesting to construct
  gauge theories using such a deformed quantum system, and analyze its consequences. In general, we expect that this will lead
  to new types of vertices between gauge fields and matter, and their coupling constants can be constrained from particle physics
  experiments. This can in turn lead to new and interesting bounds on the polymer length.

\end{document}